# A Comprehensive Review on the Advancement of Home Automation System

Md. Rawshan Habib[1], Md Abu Yusuf[2], W.M.H Nimsara Warnasuriya[3], Kumar Sunny[4], Mohammed Mahbubur Rahaman[2], Md Rezaul Karim Khan[2], Partha Pratim Saha[2], Mohammad Tanzimul Alam[4]

[1]School of Engineering and Information Technology, Murdoch University, Murdoch, Australia
[2]Department of Computer Science, Maharishi International University, IA, USA
[3]School of Engineering, Edith Cowan University, Joondalup, Australia
[4]Faculty of Computer Science, Technische Universität Chemnitz, Chemnitz, Germany
mdrawshan15@gmail.com, samba.yusuf@gmail.com, 96nimsara@gmail.com, kumarsunny1348@gmail.com, mahbubur_rahaman@outlook.com, mdrezaulkhan@gmail.com, sahapratim2@gmail.com, tanzimul.tanim@gmail.com

*Abstract*— In light of its many benefits, home automation systems are one of the subjects that are becoming ever more prevalent. The term "home automation" describes the remote monitoring and management of household equipment. The Internet and its usages are constantly expanding, which means there is a lot of room for remote access, management, and surveillance of these network-enabled systems. Nowadays, scientists and researchers are developing cutting-edge prototypes of home automation system which includes smart lighting system, smart kitchen, smart fire protection devices, smart lawn mower, smart health monitor system and so on. Each automated system's primary objective is its ability to reduce human labor, effort, time, and mistakes brought on by carelessness. The objective of this study is to provide in-depth evaluation of the newly developed home automation system. Moreover, state-of-the-art home automation topologies such as ZigBee, Z-wave, Wi-Fi and Bluetooth are also discussed here. The authors are optimistic that this study would have a major impact on the present advances in home automation technology.

## I. INTRODUCTION

Automation is becoming more and more important to everyday existence and the economy at large. Researchers work to build intricate structures for a continuously growing variety of functions and human endeavors by fusing automated technologies with quantitative and managerial abilities [1]. One of the most exciting advances in automation and control systems technology is the smart home. In 1984, the American Association of House Builders unveiled the first iteration of the smart home. This field had considerable expansion after then. The term "smart home" does not just refer to a physical structure. However, given the current state of technology, it possesses a wide connotation. The term refers to a comprehensive definition of creative or smart lifestyles [2]. Fig. 1 depicts a glimpse of smart home.

The levels of automation and intelligence in home automation differ. It might be as basic as wireless lighting management or as intricate as microcontroller-based systems. Remote connection and surveillance of household equipment and systems is the primary feature of a home automation system. When home automation systems are used, there is an integrated communication between the gadgets. Several positive aspects are obtained, including security, energy savings, and comfort. Robots, simulated representatives, and voice-controlled devices may also offer people a certain amount of company while enhancing their satisfaction of managing their homes. As a result, they possess the ability to significantly enhance physical and psychological wellness. They can also be useful in promoting self-reliance and enhancing the standard of living. Even with these possible advantages, not many households adopt smart home technology because of their exorbitant prices and labor-intensive upkeep although the fact that it might be quite helpful. Certain technologies offer alternatives which do not seem particularly practical for usage in homes [3, 4].

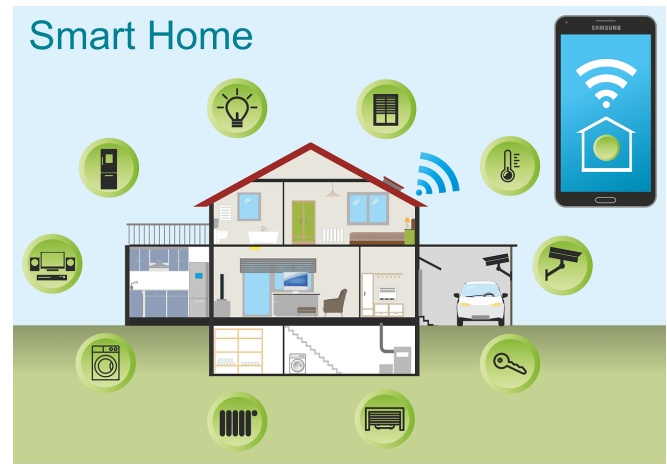

Fig. 1. Glimpse of smart home [5].

Homeowners and industry professionals have long been aware with smart homes and the theories that go along with them. But the widespread use of technology in households remains a long way off due to a lack of expertise and Specific, Measurable, Achievable, Realistic, and Timely (SMART) understanding. The most recent study offers an overview of the cutting-edge technologies utilized in smart homes, the ways they affect day-to-day activities, and the way to plan, create, build, and accomplish goals for smart homes using SMART tools [6]. In this study, advances in home automation technologies, for example, smart lighting system, smart kitchen, smart lawn mower, smart fire protection system, smart health monitoring system are discussed with some cutting-edge prototypes developed in recent years. Automation techniques, known as the foundation of the growth of home automation are also discussed in this paper with pros and cons.

## II. HOME AUTOMATION TECHNIQUES

Home automation systems used to be primarily concerned with remotely controlling loads via telephone systems and networks. Dual Tone Multi-Frequency signals were first used as an interface for home automation via computers and the telephone network. Eventually, these devices got replaced by circuits made specifically for home automation, which are composed of microcontrollers and logic circuits. Several household appliances might be effectively driven and controlled by the control circuit. A few academics created a wireless LAN-based home automation system at an ulterior point. Developers used a GSM modem to create a home automation system that utilized SMS technology when mobile communications were invented in the 1980s [7].

In the modern smart home, many techniques operate including ZigBee, Z-wave, Bluetooth, Wi-Fi, and so on. Comparison between these topologies is shown in Table I. ZigBee was initially commercially available in 2003. Its operating frequency varies depending on location, for example, it operates in 915 MHz and 868 MHz at North America and Europe, simultaneously. It is based on IEEE 802.15.4 standard. Its low power consumption, low cost and short-range coverage make it an ideal one for home automation systems. Despite of having numerous advantages, it has some drawbacks too. Orphan node problem is one of them which is addressed in [8] with a solution for network recovering.

Z-wave, low power automation technique was introduced based on ITU-T G.9959 standard. It's stable mesh topology allows it to link and regulate a variety of smart devices. Its operating frequency is 900 MHz and covers approximately 30 meters of range. It provides higher security with high cost. Short-range wireless technology known as Bluetooth is built upon the IEEE 802.15.1 standard. It offers a very affordable alternative for home automation. It supports star and point-to-point topologies. Bluetooth connections are not appropriate for large-scale smart home systems since they are usually restricted to one-to-one or one-to-many device pairings. Due to its short coverage (10 meters), more relay devices are necessary for coverage expansion. Based on the IEEE 802.11 standards, Wi-Fi is another wireless technology that is used to link equipment for home automation. It is mostly utilized in home automation for wireless appliance management and surveillance. Its operating frequency is 2.4 GHz. Wi-Fi allows fast data transfer with high power consumption [9, 10].

TABLE I. COMPARISON AMONG HOME AUTOMATION TECHNIQUES.

| Name | ZigBee | Z-wave | Bluetooth | Wi-Fi |
|---|---|---|---|---|
| Standard | IEEE 802.15.4 | ITU-T G.9959 | IEEE 802.15.1 | IEEE 802.11 |
| Frequency Band | 2.4 GHz | 900 MHz | 2.4 GHz | 2.4 GHz |
| Topology | Mesh, star, hybrid structures | Mesh | Star, point-to-point | Star |
| Range | 100 meters | 30 meters | 10 meters | 100 meters |
| Power usage | Low | Low | Low | High |
| Cost | Low | High | Very low | Medium |

III. SMART HOME APPLICATIONS

In this section, some of the cutting-edge smart home devices including smart lighting system, smart kitchen, fire protection system, smart lawn mower and smart health monitor system are discussed.

*A. Smart Lighting System*

The most prevalent embedded device in home automation is lighting control module. Both residential and commercial buildings use a significant amount of electricity for lighting. Therefore, to reduce energy usage, improved lighting systems are required. Up to 30% of power expenses can be cut with smart lighting control. An Arduino-based smart lighting system is presented in [11] where Wi-Fi topology was utilized to regulate the system. It ensures both automatic and manual control. An FPGA board was used by experts to design and test a lighting control system. When modified to meet the demands of a family living in a tiny house, the system demonstrated functionality. Users don't need to do any programming; the system operates by itself and modifies itself based on use parameters [12].

An IoT-based intelligent lighting platform is designed in [13] where ZigBee wireless network is utilized. ZigBee is being used as a communication link to lower electrical device power usage. The appliance is capable of tracking energy to save energy. System architecture is shown in Fig. 2. On the other hand, reviews are conducted in [14] on how smart LED bulbs utilized in smart lighting systems affect energy usage. The power consumption of the various colors generated by the smart LED light is also examined. It has been observed that the power consumption of the various colors that result from the smart LED bulb varies. In addition, a thorough comparison of the energy-saving capabilities of halogen, CFL, LED, and smart LED is examined. Only when a smart LED bulb is dimmed and operated remotely does it appear to have the lowest energy usage.

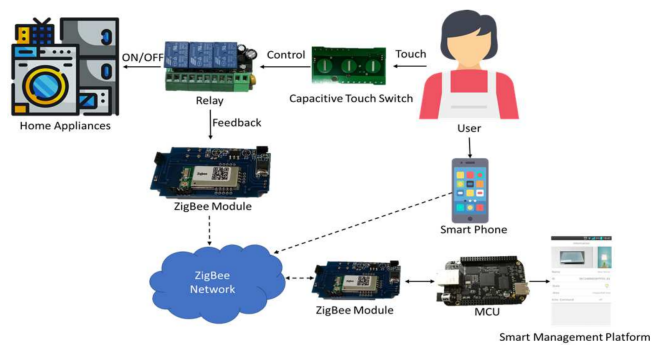

Fig. 2. Architecture of the proposed system in [13].

*B. Smart Kitchen*

The way human cooks might be completely changed by AI-controlled smart kitchen thanks to its linked equipment, automated procedures, and user-friendly interface. Cooking may be made less hazardous, more efficient, and easier with an AI-controlled kitchen that makes use of the newest machine learning technology. Meal preparation may be done with 50% less time and 70% less energy when using such technology. The smart kitchen initiative is described in [15] to offer consumers solid performance at a reasonable cost. In addition to assisting with safety, the smart kitchen notifies the user of the things they will need to purchase. On the other hand, a lot of learning may happen in the kitchen for young kids. Cooking is a fun hobby that can be shared and is excellent for fostering social connection, thought-provoking conversation, and educational enjoyment. The purpose of the E-care eating table designed in [16] is to help parents to establish friendly relationships with their kids and provide fresh environment to the kitchen. Children may learn about the shapes, colors, and weight of food products from their parents while they cook. Playing the interactive game might help them increase their vocabulary. Children may learn about plant growth and changes by planting the virtual fruits and veggies. Though the proposed table was designed with Chinese characters, it could be modified with other languages. Fig. 3 depicts virtual game interface on E-care eating table.

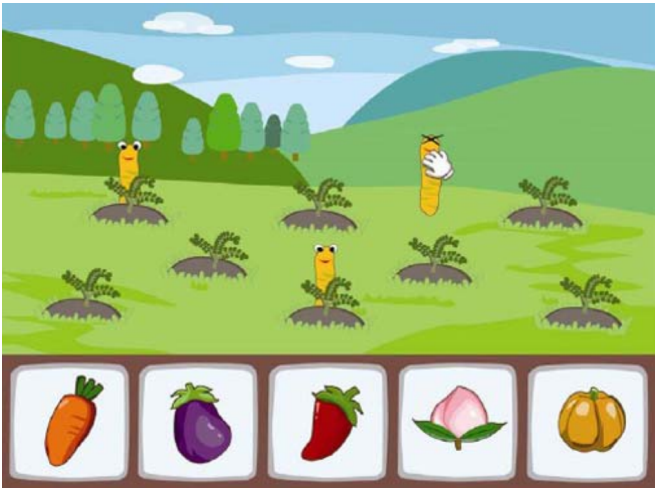

Fig. 3. Virtual game interface designed for children in [16].

Sudden explosions can occur in kitchen for a variety of causes, including uncontrolled fires, rapid temperature increases, gas leaks, etc. The explosions must be identified and removed right away. With the vision of establishing a safe kitchen, DTH11, IR, and MQ-3 sensors serve as models for an intelligent and secure monitoring system presented in [17]. Relays are used to regulate this safety system and Arduino UNOs are used to interface these sensors. This system is monitored and controlled once every fifteen seconds. The user will receive data about the kitchen via the Wi-Fi module. Applying the Internet of Things (IoT) framework, researchers created an effective smart kitchen platform in [18] which identifies gas leaks, provides update to user, closes cylinder's knob, opens the windows and starts the exhaust fan. This automated system delivers user-friendly operation using a mobile IoT interface.

*C. Fire protection System*

Fire protection is obviously important for human safety since improper handling of fire sources can result in major mishaps. The higher burning rate of contemporary house fires must be addressed to protect people lives and property. One crucial strategy for this is early fire warning systems. A fire detection and extinguishing system is created in [19] and put into practice in an artificial setting that is kept at 27°C. This system employs many functioning sensors to mitigate the risk of an alarm circuit collapse. Additionally, every sensor is used twice to increase the system's dependability. To put out the fire rapidly, this system features a fire extinguisher and a water supply. Another prototype on the planning and execution of an IoT-integrated fire alert system to facilitate early fire detection and reaction is presented in [20]. The main goal of this initiative is to improve fire safety protocols by utilizing IoT technology to identify fires early on, enabling timely intervention to reduce property damage, save lives, and facilitate rescue efforts. System model of this proposed system is depicted in Fig. 4.

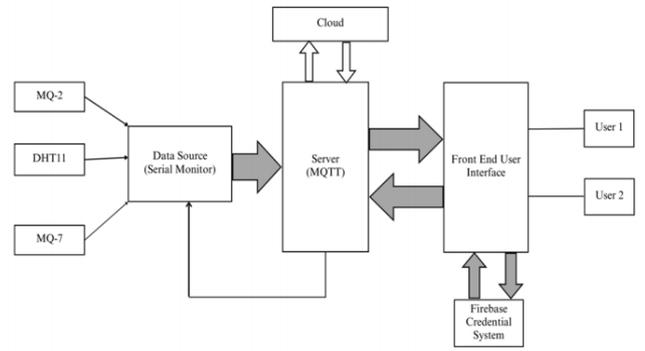

Fig. 4. Fire alert system model [20].

An interior fire early warning method based on a back propagation neural network is suggested in [21] to enhance real-time fire alarm effectiveness. According to the test findings, the suggested algorithm can accurately identify and report six standard test fires, saving 32% of the fire detection time. For smoke detection, energy conservation prototype can be utilized that is presented in [22]. This prototype showed effectiveness of opening emergency exit and entrance door in case of fire. Using a flame sensor, Raspberry Pi and Google Cloud Messaging Service (GCM), a fire detection system is developed in [23] where IoT concept is adopted to notify users of an emergency when a flame is detected so they can respond appropriately. As a result, outcome of this system aids individuals in adopting the essential safety measures for their homes.

*D. Smart Lawn Mower*

Currently available technologies for cutting grass include electric lawnmowers, internal combustion engines, as well as manually operated gadgets. These methods are labor-intensive and harmful to the environment. A unique concept is presented in [24] for a solar-powered automated machine that can be controlled by a Bluetooth module on a mobile phone. Utilizing solar panels benefits the environment by reducing pollution and supplying electricity. Another concept and implementation of an autonomous lawn cutting equipment is presented in [25] that runs mostly on solar energy. The suggested lawn cutter is affordable, lightweight, and portable. It is suggested to use two degree-of-freedom PID controllers to regulate the prototype's motor speed. Fig. 5. Presents a framework of smart lawn mower.

Applying the features of an Android-based smartphone, researchers construct a robot lawnmower and its control assistance network in [26]. The constructed robot can move manually in the following directions: forward, backward, left, and right. It can also make adjustments by responding to orders received from smartphones or go around the designated land area on its own. Naturally, with an ultrasonic sensor module added to stop the robot from colliding with solid things in front of it. A solar-powered lawnmower was developed in [27] that used a 550-motor, which produced enough torque to run the mower's solar panels. It was discovered that the solar panel's average efficiency was 93%. The pace at which power dissipates in the solar-powered lawn mower was determined by examining the battery voltage drop during the mowing of various grass varieties. mowing tough grass resulted in a voltage decrease of 0.34 V, but mowing gentle grass only caused a 0.17 V voltage drop. When rains become apparent, a rain sensor was included to notify the user using Buzzer.

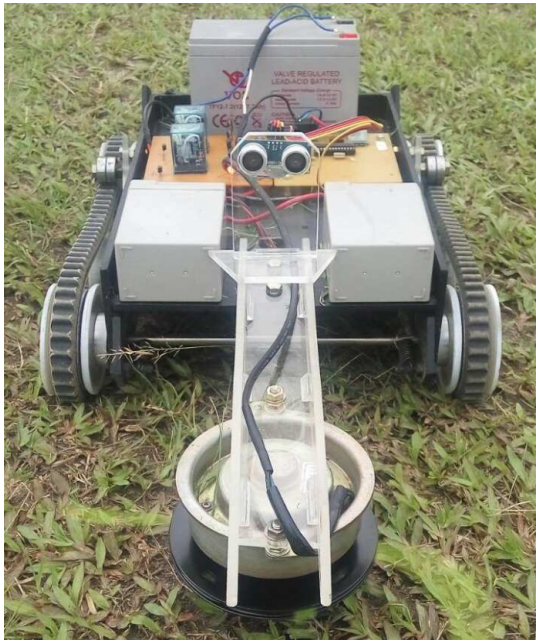

Fig. 5. Framework of smart lawn mower presented in [26].

*E. Smart Health Monitor System*

Smart home designs provide several advantages, which includes the ability to regulate the home environment by straightforward instructions and the ability to alert users to potential issues within the premises. The main beneficiaries of smart home applications are elderly individuals living alone in their homes, those with chronic illnesses, and persons with disabilities. Youthful people can benefit from a smart home setting by staying in touch with elderly relatives and being alerted to circumstances that could endanger their health [4]. In the current day, smart old age homes are recognized as an appropriate medium for elderly and disabled people to live independently and comfortably. The wireless home automation system in [28] enables detecting falls, health monitoring, position tracking, and voice control switching. Voice control switching systems allow for remote control operation of many household appliances, including TVs, fans, and lights. Clinical viewpoints and intensive care make use of the location tracking and health monitoring system. An additional fall detection system is included to prevent inadvertent harm and fatalities. A glove-based home automation system is created in [29] that can control devices automatically by identifying gestures. The design model is shown in Fig. 6. When using the glove, basic motions may be used to operate household equipment. A smartphone app is also being created so that family members may monitor the usage and status of the gadgets. As a result, the system maintains those responsible updated and offers relaxation to the special abled.

Another effective, energy conserving and economic glove-based automation system is presented in [30] where gesture recognition technique is utilized. The Arduino Mega 2560 is used as the CPU in the voice and touch aided home automation system that is developed in [31]. The system expands on the usage of currently available smartphone applications, that enable individuals to utilize Bluetooth connectivity to operate up to 44 devices, including several smoke detectors and PIR motion sensors. In contrast to previous systems, NFC tags may be used to launch and operate the complete system, making it easier to use. The entire work has been completed for less than USD 30, making it accessible to the public and maybe modernizing every single old house.

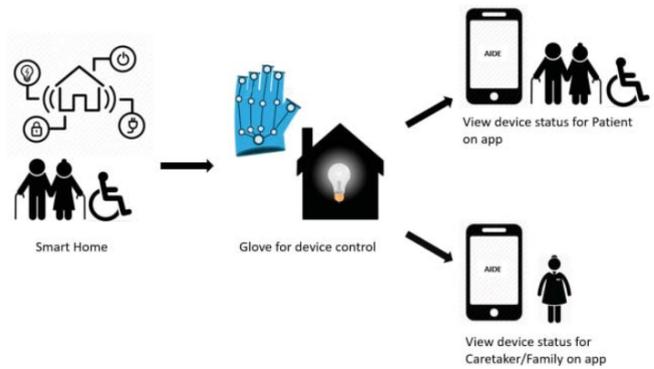

Fig. 6. Home automation for elderly and disabled people [28].

IV. CONCLUSION

Over the past several years, smart home appliances have grown in popularity and have the ability to enhance living conditions. Given the widespread usage of such innovations, it is critical to comprehend the various elements that might lead to either success or failure. Therefore, it is necessary to discuss about the latest developments of home automation system. We have discussed the latest developments in home automation technology, including smart lighting, smart kitchen, smart lawnmower, smart fire prevention, and smart health monitoring systems, along with several innovative prototypes created recently. This study also discusses the advantages and disadvantages of automation techniques, which are considered the cornerstone of the development of home automation.